\documentclass[12pt,a4paper]{article}
\usepackage{jheppub}
\pdfoutput=1
\newcommand{\be}{\begin{equation}}
\newcommand{\ee}{\end{equation}}
\newcommand{\nn}{\nonumber}

\newcommand{\LL}{\mathcal{L}}

\newcommand{\N}{\mathcal{N}}

\newcommand{\V}{\mathcal{V}}

\newcommand{\et}{\tilde{e}}

\newcommand{\p}{\partial}
\newcommand{\gat}{\tilde{\gamma}}

\newcommand{\hht}{\hat{t}}
\newcommand{\hr}{\hat{r}}
\newcommand{\hx}{\hat{x}}
\newcommand{\hy}{\hat{y}}
\newcommand{\hz}{\hat{z}}

\newcommand{\al}{\alpha}
\newcommand{\bet}{\beta}
\newcommand{\ga}{\gamma}         
\newcommand{\de}{\delta}        \newcommand{\De}{\Delta}
\newcommand{\ep}{\epsilon}
\newcommand{\ve}{\varepsilon}

\newcommand{\la}{\lambda}       \newcommand{\La}{\Lambda}

\newcommand{\si}{\sigma}

\newcommand{\om}{\omega}        

\preprint{TIFR/TH/14-09}

\title{Stable Bianchi III attractor in \boldmath{$U(1)_R$} gauged supergravity}
\author[a]{Karthik Inbasekar,}
\author[a]{Rickmoy Samanta}
\affiliation[a]{Tata Institute of Fundamental Research,\\Department of Theoretical Physics,\\ Mumbai
400005, India}
\emailAdd{ikarthik@theory.tifr.res.in}
\emailAdd{rickmoysamanta@theory.tifr.res.in}
\abstract{Bianchi attractors are homogeneous but anisotropic extremal black brane horizons. We
study the $AdS_3 \times \mathbb{H}^2$ solution which is a special case of Bianchi type III
in a $U(1)_R$ gauged supergravity. For a wide range of values for certain free parameters in gauged
supergravity, there exist a large class of solutions that satisfy conditions for the attractor
mechanism to hold. We investigate the response of the solution against linearized fluctuations of
the scalar field. The sufficient conditions for the attractor mechanism ensure that there exist a
solution for the scalar fluctuation which dies out at the horizon. Furthermore, we solve for the
gauge field and metric fluctuations that are sourced by scalar fluctuations and show that they are
well behaved near the horizon. Thus, we have an example of a stable Bianchi attractor in gauged
supergravity. We also analyze the Killing spinor equations of gauged supergravity in the
background of our solution. We find that a radial Killing spinor consistent with the Bianchi III
symmetry breaks supersymmetry.} 

\begin{document}
 \maketitle
\section{Introduction}
Recent progress in the studies of extremal black holes in Anti de-Sitter space have witnessed the
beginning of a dialogue between gravity and condensed matter physics. In gauge-gravity
duality \cite{Aharony:1999ti}, extremal solutions provide the dual gravity description of zero
temperature ground states of strongly coupled field theories. Many condensed matter theories
exhibit a wide variety of phases. In particular, systems at quantum criticality can be strongly
coupled and display novel phase transitions due to quantum fluctuations at zero temperature
\cite{sachdev2007quantum}. The subject is an active area of research and we
refer the reader to some of the review articles for references
\cite{Hartnoll:2009sz,Sachdev:2011wg,Herzog:2009xv}. 

Given such a large number of phases in condensed matter systems, it is reasonable
to expect that there is also a similar zoo of extremal solutions in the dual gravity side. Earlier
studies focused on extremal systems with translational and rotational symmetry that exhibit Lifshitz
scaling and hyperscaling violations
\cite{Goldstein:2009cv,Goldstein:2010aw,Pal:2009yp,Taylor:2008tg,Kachru:2008yh,
Balasubramanian:2008dm,Perlmutter:2010qu}. In some cases, such solutions have
been embedded in string theory
\cite{Dong:2012se,Perlmutter:2012he,Balasubramanian:2010uk,Donos:2010tu,Gregory:2010gx,
Narayan:2012hk,Dey:2012tg,Dey:2012rs}. Extremal black branes dual to field theories with reduced
symmetries are also equally interesting and have been studied 
\cite{Donos:2011qt,Donos:2011ff,Donos:2012gg,Iizuka:2012iv,Iizuka:2012pn,Donos:2012wi,
Cremonini:2012ir,
Erdmenger:2013zaa,Iizuka:2014iva}. 

Recently, new classes of extremal solutions exhibiting reduced symmetries have been found
\cite{Iizuka:2012iv,Iizuka:2012pn}. These metrics are homogeneous but anisotropic extremal black
brane horizons in five dimensions. They have been classified using the Bianchi classification
\cite{1975classical,ryan1975homogeneous}, which is well known in cosmological context and are now
known as the ``Bianchi attractors". These geometries arise as \emph{exact} solutions to gravity
coupled to simple matter in the presence of a cosmological constant. Recently, Bianchi type metrics
satisfying reasonable energy conditions have been shown to numerically interpolate to Lifshitz or
$AdS_2\times S^3$ from which they can be connected to $AdS_5$ \cite{Kachru:2013voa}. This provides
some evidence towards the expectation that they are attractor geometries.

The attractor mechanism has been thoroughly studied for extremal black holes in supergravity
theories \cite{Ferrara:1995ih,Strominger:1996kf}.\footnote{See 
\cite{Bellucci:2007ds,Ferrara:2008hwa} for recent reviews on the subject.} Originally studied for
supersymmetric black holes, it was understood later that the attractor mechanism is a consequence of
extremality rather than supersymmetry \cite{Ferrara:1997tw}, and has been shown to work for extremal
non-supersymmetric black holes \cite{Sen:2007qy,Goldstein:2005hq}. Recently much progress has been
made towards the generalization of attractor mechanism for gauged
supergravity theories
\cite{Halmagyi:2013qoa,BarischDick:2012gj,Klemm:2012yg,Barisch:2011ui,Kachru:2011ps,
Dall'Agata:2010gj,Hristov:2010eu,Ceresole:2001wi,Cacciatori:2009iz,Inbasekar:2013vra,
Inbasekar:2012sh}. The simplest Bianchi type I geometries such as Lifshitz geometries have already
been embedded in gauged supergravity \cite{Cassani:2011sv,Halmagyi:2011xh}.

A prescription fairly general enough to capture the essential features of homogeneous
geometries as generalised attractor solutions of gauged supergravity was given in
\cite{Kachru:2011ps}. The generalised attractors are defined as solutions to equation of motion
when all the fields and curvature tensors are constants in tangent space. These solutions are
characterised by constant anholonomy coefficients and are regular by construction. Following this
prescription some of the Bianchi type geometries were embedded in five dimensional
gauged supergravity \cite{Inbasekar:2012sh}. 

The generalised attractor solutions existed at critical points rather than an absolute minimum of
the attractor potential. The stability of such solutions for small perturbations of the scalar
fields about the attractor value were studied \cite{Inbasekar:2013vra}. By stability, we mean an
investigation on the response of a system subject to linearized perturbations of the fields
about their fixed point values. If the perturbations are regular as opposed to being divergent when
one approaches the fixed point, then it is a stable attractor. There is also the notion of stability
as described by the B.F. bound \cite{Breitenlohner:1982jf,Breitenlohner:1982bm}. However, we do not
discuss this here. 

It was found in \cite{Inbasekar:2013vra}, that the stress energy tensor in gauged supergravity
depends on linearized scalar fluctuations due to the interaction terms. Therefore, for back-reaction
to be small as one approaches the attractor geometry, the scalar
fluctuations are required to be regular near the horizon. For the solutions constructed in
\cite{Inbasekar:2013vra,Inbasekar:2012sh}, the scalar fluctuations about the critical values were
regular near the horizon only when the Bianchi geometries factorized as $AdS_2 \times M$, where M is
a homogeneous space of dimension three. The factorized geometries have the unphysical property that
the entropy does not vanish as the temperature goes to zero.

In this work, we seek to study an interesting class of Bianchi type solutions which do not
factorize and are stable under linearized scalar fluctuations. Our strategy is to rely on the
conventional  wisdom  of the physics of stable attractor points for extremal black holes. Namely,
there are two sufficient conditions for the attractor mechanism
\cite{Goldstein:2005hq}. First, there must exist a critical point of the effective potential.
Second, the Hessian of the effective potential evaluated at the solution must have positive
eigenvalues. These two conditions are always met by supersymmetric solutions. For non-supersymmetric
extremal black hole solutions the above two conditions are sufficient to guarantee a stable
attractor.

Keeping the above strategy in mind, we study the $AdS_3 \times \mathbb{H}^2$ solution which is a
special case of Bianchi type III in gauged supergravity. Supersymmetric $AdS_3 \times
\mathbb{H}^2$ solutions have been studied earlier in $U(1)^3$ gauged supergravity
\cite{Klemm:2000nj}. In the context of wrapped branes, $AdS_3 \times \mathbb{H}^2$ solutions have
been constructed in type IIB supergravity compactified on $S^5$ \cite{Maldacena:2000mw}. We consider
the $U(1)_R$ gauged supergravity \cite{Gunaydin:1999zx,Gunaydin:2000xk} for our study. We find that
there are a large class of type III solutions that exist at a critical point corresponding to a
minimum of the attractor potential. We do a linearized fluctuation analysis of the scalar field
about its attractor value. For the scalar fluctuations sufficient conditions for a stable attractor
guarantees the existence of a solution which dies out at the horizon. We then determine the gauge
field and metric fluctuations that are sourced by scalar fluctuations. We find that the simplicity
of the solution causes the source term in the gauge field fluctuations to vanish. Hence there are no
gauge field fluctuations sourced by the scalar fluctuations in this case. As a result the metric
fluctuations are sourced purely by scalar fluctuations. We solve the equations for the metric
fluctuations with the source terms and show that they vanish as one approaches the horizon. 

The results of the stability analysis are as follows. The Bianchi type III metric
\be
ds^2= -\hr^{2 \beta_t}d\hht^2+ \frac{d \hr^2}{\hr^2}+ d\hx^2+ e^{-2\hx}d\hy^2+
{\hr}^{2\beta_t}d\hz^2 
\ee
which has the scaling symmetries
\be
\hht\rightarrow \frac{\hht}{\al^{\beta_t}} \ , \quad \hr\rightarrow \al \hr \ , \quad
\hx\rightarrow \hx \ ,\quad \hy\rightarrow \hy \ , \quad \hz \rightarrow \frac{\hz}{\al^{\beta_t}}
\ ,
\ee
is a generalised attractor solution in gauged supergravity. The solution exists at a critical point
$\phi_c$ such that
\be
\frac{\p\V_{attr}}{\p\phi}\bigg|_{\phi_c}=0 \ , \quad
\frac{\p^2\V_{attr}}{\p\phi^2}\bigg|_{\phi_c}>0 \ ,
\ee
where $\V_{attr}$ is the attractor potential. The above conditions are expressed in terms of some
free parameters in gauged supergravity that are not fixed by any symmetries and are met for a wide
range of values. Thus a class of solutions exists at a minimum of the attractor potential and the
Hessian has a positive eigenvalue. The scalar field fluctuations $\de\phi$ about the attractor
values are of the form
\be
\de\phi\sim \hr^\De \ , \quad \De>0 \ .
\ee
The scalar fluctuations are regular near the horizon $\hr\rightarrow 0$. All the metric fluctuations
$\ga_{\mu\nu}$ are of the form
\be
\ga_{\mu\nu}\sim g_{\mu\nu} \hr^\De
\ee
and are regular near the horizon. Thus, we have a class of Bianchi III solutions which are stable
with respect to linearized fluctuations of scalar, gauge field and metric fluctuations about the
attractor value. The solution is an example of a stable Bianchi attractor in gauged
supergravity.

Given that the solution is a stable Bianchi attractor, we also investigate its supersymmetry
properties. The study of supersymmetry of Bianchi attractors is very interesting since it can lead
to solutions such as domain walls interpolating between Bianchi attractors and $AdS$.
Besides, supersymmetry equations are first order differential equations and are often easier to
solve. Earlier studies on supersymmetry of Bianchi type metrics have focused on the Bianchi I
class. The simplest of which is $AdS$ space. In this case, there are two types of Killing spinors,
one which is purely radial and the other which depends on all coordinates
\cite{Lu:1996rhb,Lu:1998nu}. The radial spinor generates the Poincar\'{e} supersymmetries
while the other spinor generates the conformal supersymmetries. The earliest works were
on supersymmetric black string solutions whose near horizon geometries take the form $AdS_3 \times
\mathbb{H}^2$ \cite{Cacciatori:2003kv,Klemm:2000nj}. The Supersymmetry of the Bianchi I metrics such
as Lifshitz, have also been studied in four dimensional gauged supergravity
\cite{Cassani:2011sv,Halmagyi:2011xh}. In five dimensional $U(1)^3$ gauged supergravity Bianchi I
types such as $AdS_2\times \mathbb{R}^3, AdS_3\times\mathbb{R}^2$ have been found to be
supersymmetric \cite{Almuhairi:2011ws}. In the above cases the geometries preserve $1/4$ of the
supersymmetry and the Killing spinor equations were solved for a spinor which depended only on the
radial direction.

In this spirit, we study the Killing spinor equations of $\N=2 , U(1)_R$ gauged supergravity in the
background of the Bianchi type III metric. We choose the radial ansatz for the Killing
spinor, since it preserves the time translation symmetries and homogeneous symmetries of
the type III metric. However, we find that the radial ansatz breaks all the supersymmetries. This
suggests that the stable type III solution that we have constructed may be a non-supersymmetric
attractor.

The paper is organised as follows. In \S\ref{Bianchi3inEM} we construct a magnetic Bianchi type III
solution in Einstein-Maxwell theory with massless gauge fields. Following that, we provide some
background in $U(1)_R$ gauged supergravity and generalised attractors in \S\ref{gaugedsugra} and
\S\ref{genatt}. In the next subsection \S\ref{typeIIIsoln} we embed the Bianchi type III solution
in the $U(1)_R$ gauged supergravity. We discuss the linearized fluctuation analysis of the gauge
field, scalar field and metric in \S\ref{linearisedflucabtattr}. We analyze the Killing spinor
equation in gauged supergravity with the background Bianchi type III metric in \S\ref{susy}. We
conclude and summarize our results in \S\ref{summary}. We summarize some of the notations and
conventions in \S\ref{notations}. We provide some details regarding the linearized Einstein
equations in \S\ref{linearisedeeq} and list the coefficients that appear in the metric
fluctuations in \S\ref{funcofbetat}.

\section{Bianchi III solution in Einstein-Maxwell theory}\label{Bianchi3inEM}

We begin with a quick review of some elements of the Bianchi III symmetry. The Bianchi
classification of real Lie algebras in three dimensions is well known in the literature
\cite{ryan1975homogeneous,1975classical}. There are nine types of such algebras. In three
dimensional Euclidean space, Killing vectors that generate homogeneous symmetries close to form Lie
algebras that are isomorphic to the Bianchi classification.

The Bianchi III algebra is generated by the Killing vectors
$X_i$
\be\label{killingvec}
X_1=\p_{\hy} \ , \quad X_2=\p_{\hz} \ ,\quad X_3=\p_{\hx}+\hy \p_{\hy}\ ,
\ee
\be
[X_1,X_3 ]= X_1 \ .
\ee
The only non trivial Killing vector is the translation in the $\hx$ direction that is
accompanied by a unit weight scaling in the $\hy$ direction. To write a metric which is
manifestly invariant under this symmetry, one identifies the vector fields $\et_i$ that commute with
the Killing vectors 
\be
[\et_i, X_j]=0 \ .
\ee
The invariant vector fields for the type III case are
\be
\et_1 = e^{\hx} \p_{\hy} \ , \quad \et_2 =\p_{\hz} \ , \quad \et_3 =\p_{\hx} \ ,
\ee
\be\label{algebra}
[\et_1,\et_3]=-\et_1 \ , [\et_1,\et_2]=0 \ , [\et_2,\et_3]=0 \ .
\ee
Note that $\et_1$ and $\et_3$ form a sub-algebra. This sub-algebra is generated by the Killing
vectors of the hyperbolic space $\mathbb{H}^2$ in two dimensions. The two dimensional analogue of
the Bianchi classification consists of two distinct algebras. One is a trivial algebra with
commuting generators corresponding to $\mathbb{R}^2$ and the other is the algebra that corresponds
to $\mathbb{H}^2$ \cite{1975classical}. 

The duals of the $\et_i$ are one forms $\om^i$
\be\label{invariantoneforms}
\om^1= e^{-\hx}d\hy \ ,\quad \om^2=d\hz \ ,\quad \om^3= d\hx \ ,
\ee
that are invariant under the type III homogeneous symmetry. The invariant one forms satisfy the
relation
\be
d\om^1= \om^1\wedge\om^3 \ .
\ee
The metric written in terms of the invariant one forms
\be
ds_3^2= (\om^1)^2+(\om^2)^2+(\om^3)^2
\ee
is manifestly invariant under the homogeneous type III symmetries.

We are interested in five dimensional black brane horizons with homogeneous symmetries in
the spatial directions. These geometries are obtained from gravity coupled to simple matter in
the presence of a cosmological constant and are known as the Bianchi attractors
\cite{Iizuka:2012iv,Iizuka:2012pn}. For the purposes of this article, we construct a simple type III
solution in Einstein-Maxwell theory sourced by a single massless gauge field and a cosmological
constant. We take the type III metric to be of the form
\be\label{typeIII}
ds^2= -\hr^{2 \beta_t}d\hht^2+ \frac{d \hr^2}{\hr^2}+ (\om^3)^2+ (\om^1)^2+ {\hr}^{2\beta_2}
(\om^2)^2 \ ,
\ee
where $\beta_t, \beta_2$ are positive exponents. For the case $\beta_t=\beta_2$, the metric becomes
$AdS_3 \times EAdS_2$. To see this we substitute for the invariant one forms from
\eqref{invariantoneforms} and make the coordinate transformation $\hx=\ln \hat{\rho}$ to get,
\be\label{AdS2EAdS2}
ds^2=\bigl(-\hr^{2\beta_t} d\hht^2+\frac{d\hr^2}{\hr^2}+\hr^{2\beta_t}d\hz^2\bigr) +
\biggl(\frac{d\hy^2+d\hat{\rho}^2}{\hat{\rho}^2}\biggr) \ .
\ee
When one performs a Kaluza-Klein reduction of the above solution one gets the $AdS_2\times
EAdS_2$ solution in four dimensions with hyper scale violation \cite{Iizuka:2012pn}.

We now construct the Type III solution \eqref{typeIII} in Einstein-Maxwell theory. The action is of
the form 
\be
S=\int d^5x \sqrt{-g}(R-\frac{1}{4}F^{\mu\nu}F_{\mu\nu}+ \La) \ ,
\ee
where $\La>0$ corresponds to Anti de-Sitter space in our conventions.  We are interested in a
magnetic solution and we choose the gauge field to have components along the $\om^1$ direction
\be
A=A_3 \om^1 ,
\ee
where $A_3$ is a constant.\footnote{The notation $A_3$ is just chosen for convenience.} The
gauge field equations are automatically satisfied with this ansatz
and the independent trace reversed Einstein equations are
\begin{align}
A_3^2-6 \beta_t(\beta_2+\beta_t)+2 \La=0 \ , \nn \\
A_3^2-6 (\beta_2^2+\beta_t^2)+2 \La=0 \ , \nn \\
-A_3^2-3+\La=0 \ , \nn\\
A_3^2-6\beta_2(\beta_2+\beta_t)+2\La=0 \ .
\end{align}
The $\hht\hht$ and $\hz\hz$ equations imply
\be
\beta_2=\beta_t 
\ee
and the rest of the equations give the solution
\be
\La=1+4\beta_t^2 \ , \quad A_3=\sqrt{-2+4 \beta_t^2} \ .
\ee
Thus we have a magnetic type III solution sourced by a massless gauge field and parametrized by 
$\beta_t$, which satisfies the condition 
\be\label{condbetat}
\beta_t^2>\frac{1}{2} \ ,
\ee
such that $A_3$ is real. In the following section, we construct a similar solution in $U(1)_R$
gauged supergravity.
\section{Gauged supergravity and generalised attractors}\label{gaugedsugrandgenatt}
\subsection{Gauged supergravity}\label{gaugedsugra}
In this section, we review essential material in $\N=2, d=5$ gauged supergravity relevant for our
purpose. The general supergravity coupled to vector, tensor, hyper multiplets with a gauging of
the symmetries of the scalar manifold and R symmetry is discussed in
\cite{Ceresole:2000jd}. We work with the $\N=2,d=5$ gauged supergravity coupled to a single
vector multiplet and a gauging of the $U(1)_R$
symmetry \cite{Gunaydin1985573,Gunaydin1984244,Gunaydin:1999zx,Gunaydin:2000xk}. 

The gravity multiplet consists of two gravitinos $\psi_\mu^i$, $i=1,2$, and a graviphoton. The
vector multiplet consists of a vector $A_\mu$, a real scalar $\phi$ and the gaugini $\la_i$. The
vector in the vector multiplet and the graviphoton are collectively represented by $A^I_\mu$,
$I=0,1$.

The scalars in the theory parametrize a very special manifold described by the cubic surface (see
for eg \cite{deWit:1991nm})
\be\label{scalarmanifold}
N\equiv C_{IJK}h^I h^J h^k=1 \ ,\quad h^I\equiv h^I(\phi) \ .
\ee
The constants $C_{IJK}$ are real and symmetric. The condition \eqref{scalarmanifold} is solved
by going to a basis \cite{Gunaydin1985573,Gunaydin1984244},
with $h^{I}=\sqrt{\frac{2}{3}}\xi^{I}|_{N=1}$ such that,
\be
N(\xi)=\sqrt{2} \xi^0 (\xi^1)^2=1 \ ,
\ee
where,
\be
\xi^0= \frac{1}{\sqrt{2}\phi^2} ,\ \ \xi^1=\phi \ .
\ee
From the definition of the basis, we find that the $h^I$ are related to the scalars $\phi$ in the
Lagrangian through
\be\label{scalarpar}
h^{0}=\frac{1}{\sqrt{3}\phi^2}, \ \ 
h^{1}=\sqrt{\frac{2}{3}}\phi \ .
\ee
It is clear from the scalar parametrization that the only non-zero coefficients for $C_{IJK}$ are
$C_{011}=\sqrt{3}/2$ and its permutations. 

The ambient metric used to raise and lower the index $I$ is defined through 
\be\label{ambient}
a_{IJ}=-\frac{1}{2}\frac{\p}{\p h^I}\frac{\p}{\p h^J} \ln N|_{N=1} \ ,
\ee
and takes the form
\be
a_{IJ}=\begin{bmatrix}
 \phi^4 & 0 \\
0 &  \frac{1}{\phi^2}\\
\end{bmatrix} \ .
\ee
The metric on the scalar manifold is obtained from the ambient metric \eqref{ambient}
through 
\be
g_{xy}=h^I_x h^J_y a_{IJ}\ , \quad h^I_x=-\frac{\sqrt{3}}{2}\frac{\p h^I}{\p
\phi^x} \ .
\ee
Since we only have a single scalar field, using the equations \eqref{scalarpar} and
\eqref{ambient} we obtain
\be\label{scalarmetric}
g(\phi)=\frac{3}{\phi^2} \ .
\ee

The field content and the various definitions above are identical to the ungauged theory. The
difference in the gauged theory is the presence of a scalar potential. The process of gauging 
converts some of the global symmetries of the Lagrangian into local symmetries. One of the global
symmetries enjoyed by the fermions in a $\N=2$ theory is the $SU(2)_R$ symmetry. For the case of
interest, we consider the gauging of the abelian $U(1)_R\subset SU(2)_R$. The R symmetry is gauged
by replacing the usual Lorentz covariant derivative acting on the fermions with $U(1)_R$ gauge
covariant derivative as follows
\begin{align}
& \nabla_\mu\la^i\rightarrow \nabla_\mu\la^i+g_R A_\mu(U(1)_R) \de^{ij}\la_j \ , \nn \\
& \nabla_\mu\psi^i_\nu\rightarrow \nabla_\mu\psi^i_\nu+g_R A_\mu(U(1)_R)\de^{ij}\psi_{\nu j} \ .
\end{align}
We refer the reader to \S\ref{notations} for conventions on raising and lowering of the $SU(2)$
indices. The $\de_{ij}$ in the covariant derivatives are the
usual Kronecker delta symbols and $g_R$ is the $U(1)_R$ gauge coupling constant. 
The $U(1)_R$ gauge field is a linear combination of the gauge fields in the theory 
\be
A_\mu(U(1)_R)= V_I A^I_\mu \ ,
\ee
where the parameters $V_I\in \mathbb{R}$ are free.\footnote{When the
gauging of R symmetry is accompanied by gauging of a non-abelian symmetry group $K$ of the scalar
manifold, the $V_I$ are constrained by $f_{JK}^I V_I=0$, where $f^I_{JK}$ are structure constants of
$K$.}

The $U(1)_R$ covariantization breaks the supersymmetry and therefore compensating terms are added
to the Lagrangian for supersymmetric closure \cite{Gunaydin:2000xk}. These terms result in the form
of a potential for the scalar fields, 
\be\label{scalarpotentialinmodel}
\V(\phi)= -2 g_R^2 V_1 \bigg[ \frac{2\sqrt{2} V_0}{\phi} + \phi^2 V_1\bigg] \ .
\ee
The potential has a critical point at 
\be
\phi_*=\bigg(\sqrt{2}\frac{V_0}{V_1}\bigg)^{1/3} \ .
\ee
The vacuum solution at this critical point is a supersymmetric Anti de-Sitter space with a
cosmological constant $\V(\phi_*)=-6 g_R^2 V_1^2\phi_*^2 $.

The bosonic part of the Lagrangian is
\begin{align}\label{lagrangian}
\hat{e}^{-1} \LL= & -\frac{1}{2}R -\frac{1}{4} a_{I J}F_{\mu\nu}^I F^{J
\mu\nu}-\frac{1}{2}g(\phi)\p_\mu
\phi\p^\mu\phi\nn\\
&-\V(\phi)+\frac{\hat{e}^{-1}}{6\sqrt{6}}C_{IJK}\epsilon^{\mu\nu\rho\sigma\tau}
F^I_{\mu\nu}F^J_{\rho\sigma}A^K_\tau \ ,
\end{align}
where $\hat{e}=\sqrt{-det g_{\mu\nu}}$ and $C_{IJK}$ are the constant symmetric coefficients that
appeared in the definition of the scalar manifold \eqref{scalarmanifold}. 

We also list the various field equations for reference. The gauge field equations are
\be\label{gfeq}
\p_\mu(\hat{e}a_{IJ}F^{J\mu\nu})=-\frac{1}{2\sqrt{6}}\ep^{\nu\la\rho\si\tau}F^J_{\la\rho}F^K_{
\si\tau} \ .
\ee
The scalar field equations are
\be\label{sceq}
\frac{1}{\hat{e}}\p_\mu(\hat{e}g(\phi)\p^\mu\phi)-\frac{1}{2}\frac{\p
g(\phi)}{\p\phi}\p_\mu\phi\p^\mu\phi-\frac{\p}{\p\phi}\bigg[\frac{1}{4}a_{IJ}F^I_{\mu\nu}F^{J\mu\nu}
+\V(\phi) \bigg]=0 \ 
\ee
and the Einstein equations are
\be\label{eeq}
R_{\mu\nu}-\frac{1}{2}R g_{\mu\nu}=T_{\mu\nu} \ , 
\ee
where the stress energy tensor is
\be\label{stress}
T_{\mu\nu}=g_{\mu\nu}\big[\frac{1}{4}a_{IJ}F^I_{\mu\nu}F^{J\mu\nu}+\V(\phi)
+\frac{1}{2}g(\phi)\p_\mu\phi
\p^\mu\phi\big]-\big[a_{IJ}F^I_{\mu\la}F^{J \ \la}_{\nu}+g(\phi)\p_\mu\phi\p_\nu\phi\big] \ .
\ee

\subsection{Generalised attractors}\label{genatt}
We now outline a brief discussion on a class of solutions to the field equations
known as generalised attractors \cite{Kachru:2011ps}. For a $\N=2, d=5$ gauged supergravity with
generic gauging of scalar manifolds and in the presence of hyper/tensor multiplets, the generalised
attractor equations were shown to be algebraic in \cite{Inbasekar:2012sh}. The $U(1)_R$ gauged
supergravity discussed in \S\ref{gaugedsugra} is a special case of the general gauged theory. The
relevant field equations which follow from \eqref{lagrangian} can be simply obtained by setting the
tensors, hyperscalars and the coupling constant associated with gauging of the scalar manifold to
zero in the field equations derived in \cite{Inbasekar:2012sh}.

Generalised attractors are defined as solutions to equations of motion that reduce to algebraic
equations when all the fields and Riemann tensor components are constants in tangent space 
\be\label{genattansatz}
\phi=const\ , \ \quad A^I_a=const \ , \quad c_{ab}^{\ \ c}=const \ ,
\ee
where $a=0,1,\ldots,4$, are tangent space indices. The $c_{ab}^{\ \ c}$, referred to as anholonomy
coefficients are structure constants that appear in the Lie bracket of the vielbeins 
\be
[e_a,e_b]=c_{ab}^{\ \ c}e_c\ ,\quad e_a \equiv e_a^\mu\p_\mu \ .
\ee
In the absence of torsion, the spin connections are expressed in terms of the
anholonomy coefficients
\be
\om_{abc}=\frac{1}{2}(c_{abc}-c_{acb}-c_{bca}) \ ,
\ee
which are constants.\footnote{The antisymmetry properties of the spin connection and anholonomy
coefficients are $\om_a^{\ bc}=-\om_a^{\ cb}$ and $c_{ab}^{ \ \ c}=-c_{ba}^{\ \ c}$ respectively.}
Thus the curvature tensor components expressed in
terms of the spin connections as 
\be
R_{abc}^{\ \ \ d}=-\om_{ac}^{\ \ e}\om_{be}^{\ \ d}+\om_{bc}^{ \ \
e}\om_{ae}^{\ \ d}-c_{ab}^{\ \ e} \om_{ec}^{\ \ d}
\ee
 are constants in tangent space. Hence, the generalised attractor solutions characterised by
constant anholonomy coefficients and are regular. 

At the attractor points defined by \eqref{genattansatz} the scalar field equation \eqref{sceq}
reduces to the  condition 
\be\label{scattr}
\frac{\p\V_{attr}(\phi,A)}{\p\phi}=0 \ 
\ee
on an attractor potential 
\be
\V_{attr}(\phi,A)=\frac{1}{4}a_{IJ}F^I_{\mu\nu}F^{J\mu\nu}+\V(\phi) \ .
\ee
Solving \eqref{scattr} gives the critical value of the scalar $\phi_c$ in terms of the charges $A$.
The critical point is a minimum when the Hessian has positive eigenvalues, which is also the
condition for a stable attractor solution \cite{Goldstein:2005hq}.

We also list the tangent space generalised attractor equations for the gauge and Einstein equations
for reference. The gauge field equations are
\be\label{gsgfeq}
a_{IJ} (\om_{a \ c}^{ \ a} F^{J bc}+\om_{a \ c}^{ \ b}F^{J ac})=0 \ ,
\ee
where the the field strength is
\be
F^I_{ab}\equiv e_b^\mu e_a^\nu (\p_\mu e_\nu^c-\p_\nu e_\mu^c)A^I_c=c_{ab}^c A^I_c \ ,
\ee
and the Chern-Simons term vanishes for the Bianchi attractors \cite{Inbasekar:2012sh}.
The Einstein equations are
\be\label{gseeq}
R_{ab}-\frac{1}{2}R \eta_{ab}=T^{attr}_{ab} \ ,
\ee
where
\be
T^{attr}_{ab}=\V_{attr}(\phi,A)\eta_{ab}-a_{IJ}F^I_{ac}F_{\ b}^{J c} \ .
\ee
In the following section we solve the algebraic attractor equations and find a Bianchi type III
solution.
\subsection{Bianchi III solution in $U(1)_R$ gauged supergravity}\label{typeIIIsoln}
We choose the Bianchi type III ansatz as before in eq.\eqref{typeIII}. The gauge field ansatz is
also same as before,
\be\label{gaugeansatz}
A_{\hy}^I= e^{-x}A_3^I \ , \quad A_3^0 \equiv A_3 \ ,
\ee
where we have turned on only the graviphoton $I=0$ for simplicity. Similar to the Einstein-Maxwell
case studied in \S \ref{Bianchi3inEM} earlier, the gauge field equations \eqref{gsgfeq} are
trivially satisfied in the $U(1)_R$ gauged supergravity as expected. 

At the attractor point the scalars are constant. Hence the scalar equations reduce to extremization
of the attractor potential \eqref{scattr}. The attractor potential has the form
\be\label{attractorpot}
\V_{attr}(\phi,A)=\frac{1}{2\phi}\big(A_3^2 \phi^5-4 g_R^2 V_1 (2\sqrt{2}V_0+V_1\phi^3)\big)\ .
\ee
The second term is the contribution of the potential \eqref{scalarpotentialinmodel}. We would like
to briefly contrast the nature of the possible critical points possible from \eqref{attractorpot}
as compared to some of the earlier works \cite{Inbasekar:2012sh,Inbasekar:2013vra}. The Bianchi
attractors constructed in gauged supergravity  were attractor solutions such that the critical
points of the attractor potential coincided with the critical points of the scalar
potential \eqref{scalarpotentialinmodel}. This was a simplification which was possible because the
attractor potential had additional terms due to gauging of the scalar manifold or with multiple
field strengths in the absence of such gauging. For the $U(1)_R$ case with just one gauge field
considered here, the attractor potential \eqref{attractorpot} does not allow such critical points
for non-trivial gauge fields. It is also important to note that in \cite{Inbasekar:2012sh}, the
Bianchi III solution could not be obtained from the Bianchi VI$_h$ solution by taking the limit
$h\rightarrow 0$ since it resulted in a singular gauge field.\footnote{The Bianchi VI$_h$ algebra
has a free parameter $h$. The Bianchi V algebra is obtained in the limit $h\rightarrow 1$, while
the Bianchi III algebra is obtained in the limit $h\rightarrow 0$
\cite{1975classical,ryan1975homogeneous}.} 

The scalar field equation then reduces to,
\be\label{gssceq}
\frac{\p\V_{attr}(\phi,A)}{\p\phi}=\frac{2}{\phi^2}\big(A_3^2\phi^5+4 g_R^2 V_1 (\sqrt{2}V_0-V_1
\phi^3 )\big)=0 \ .
\ee
In principle, one can solve for $\phi$ from the above equation. In practice, it is much easier to
solve the scalar equation simultaneously with the Einstein equation to get nice compact expressions.

The independent Einstein equations \eqref{eeq} are 
\begin{align}\label{gseeq1}
 2(1+\beta_2^2)\phi+A_3^2 \phi^5-4 g_R^2 V_1(2\sqrt2 V_0 +V_1 \phi^3)=0\nn \ ,\\
2(1+\beta_2\beta_t)\phi+A_3^2\phi^5-4 g_R^2 V_1(2\sqrt2 V_0 +V_1 \phi^3)=0\nn \ ,\\
2(\beta_2^2+\beta_2\beta_t+\beta_t^2)\phi-A_3^2\phi^5-4 g_R^2 V_1(2\sqrt2 V_0 +V_1 \phi^3)=0\nn \
,\\
2(1+\beta_t^2)\phi+A_3^2\phi^5-4 g_R^2 V_1(2\sqrt2 V_0 +V_1 \phi^3)=0 \ .
\end{align}
From the $\hat{t}\hat{t}$ and the $\hz\hz$ equations we get
\be
\beta_2=\beta_t \ .
\ee
The equations now simplify to
\begin{align}
 2(1+\beta_t^2)\phi+A_3^2 \phi^5-4 g_R^2 V_1(2\sqrt2 V_0 +V_1 \phi^3)=0\nn \ ,\\
6\beta_t^2\phi-A_3^2\phi^5-4 g_R^2 V_1(2\sqrt2 V_0 +V_1 \phi^3)=0 \ .
\end{align}
We solve for $A_3$ from the above equations to obtain
\be\label{gaugefieldsoln}
A_3=\frac{\sqrt{-1+2\beta_t^2}}{\phi^2} \ ,
\ee
and
\be
(1+4\beta_2^2)\phi-4 g_R^2 V_1 (2\sqrt{2}V_0+V_1\phi^3)=0 \ .
\ee
This equation can be solved together with the scalar equation \eqref{gssceq} to determine the
critical point
\be\label{critpoint}
\phi_c=4\sqrt{2} g_R^2 V_0 V_1\ , \quad \beta_t=\frac{1}{2}\sqrt{1+128 g_R^6 V_0^2 V_1^4}
\ee
For the gauge field to be real we require 
\be
\beta_t^2>\frac{1}{2} .
\ee
We note that the same condition was obtained for the Type III solution in Einstein-Maxwell theory
\eqref{condbetat}. It is also clear from \eqref{critpoint} that the condition is satisfied for
arbitrary values of the gauged supergravity parameters $g_R,V_0,V_1$. 

We now examine the nature of the critical point given by eqs.\eqref{critpoint} and
\eqref{gaugefieldsoln}. The Hessian evaluated at the critical point
\be\label{hessianeigenvalue}
\frac{\p^2\V_{attr}(\phi,A)}{\p\phi^2}\bigg|_{\phi_c}=\frac{-7+8\beta_t^2}{\phi_c^2} 
\ee
is positive provided we choose
\be
\beta_t^2>\frac{7}{8} \ .
\ee
We choose this condition for $\beta_t^2$, since above this bound we also satisfy the general
condition for a stable attractor solution. In terms of the gauged supergravity parameters the
condition on $\beta_t^2$ translates to
\be\label{condgaugedsugraparameters}
g_R^6V_0^2V_1^4>\frac{5}{256} \ ,
\ee
which can be satisfied for a wide range of values for the parameters $g_R,V_0,V_1$, since none of
them are constrained in anyway. Thus, for various values of $g_R,V_0,V_1$ satisfying
\eqref{condgaugedsugraparameters} we find a class of type III Bianchi metrics as generalised
attractor solutions in $U(1)_R$ gauged supergravity.

The attractor potential evaluated at the critical point
given by \eqref{gaugefieldsoln} and \eqref{critpoint} takes a remarkably simple form 
\be\label{attrpotatcrit}
\V_{attr}|_{\phi_c}=-(1+\beta_t^2)\ ,
\ee
which will be useful later.
To summarize, the type III solution is
\begin{align}\label{typeIIIsol}
 & ds^2= -\hr^{2 \beta_t}d\hht^2+ \frac{d \hr^2}{\hr^2}+ (\om^3)^2+ (\om^1)^2+ {\hr}^{2\beta_2}
(\om^2)^2 \ , \nn\\
& A_3=\frac{\sqrt{-1+2\beta_t^2}}{\phi_c^2}, \quad \phi_c=4\sqrt{2}g_R^2 V_0 V_1 , \nn \\
& \beta_2=\beta_t, \quad \beta_t=\frac{1}{2}\sqrt{1+128 g_R^6 V_0^2 V_1^4},\quad
\beta_t^2>\frac{7}{8} \ .
\end{align}
We have seen that the Hessian of the effective potential evaluated on this solution has a positive
eigenvalue suggesting that it is a stable attractor. In the following section we provide more
evidence by considering linearized fluctuations of the scalar, gauge and metric fields about their
attractor values and showing that they are well behaved near the horizon.
\section{Linearized fluctuations about attractor value}\label{linearisedflucabtattr}
In this section, we study the linearized fluctuations of the gauge field, scalar field and metric
about their attractor values. For $\N=2,d=5$ gauged supergravity coupled to vector multiplets with
a generic gauging of the scalar manifold and gauging of R symmetry the linearized equations
were derived in \cite{Inbasekar:2013vra}. The corresponding equations for the $U(1)_R$ case that
follow from \eqref{lagrangian} can be simply obtained by setting the coupling
constant associated with gauging of the scalar manifold to zero. 

The linearized fluctuations about the attractor values are of the following form,
\begin{align}\label{linearisedfluc}
& \phi_c+ \ep\de\phi(\hr) \ , \nn\\
& A_\mu+\ep\de A_\mu(\hr) \ ,\nn\\
& g_{\mu\nu}+\ep \ga_{\mu\nu}(\hr) \ ,
\end{align}
where $\ep < 1$. The attractor values of the scalar field and gauge field are $\phi_c$, $A_\mu,$
respectively. We take the near horizon metric $g_{\mu\nu}$ as the type III Bianchi metric
\eqref{typeIIIsol}. We have chosen all the fluctuations to depend purely on the radial
direction $\hr$, since it is this behavior that is most interesting from the point of view of an RG
flow. Also, this is the first thing to attempt before going to much complicated cases. The magnetic
type III solution \eqref{typeIIIsol} offers lot of simplifications. In particular, we will see that
the source term in the gauge field fluctuations vanishes and this simplifies the procedure of
solving for the metric fluctuations later on.

\subsection{Gauge field fluctuations}\label{gaugeflucanalysis}
The equation satisfied by the linearized gauge field fluctuations is
\be\label{gaugefluc}
a_{IJ}|_{\phi_c}\nabla_\mu F^{\mu\nu J}_{\de}=-\frac{\p
a_{IJ}}{\p\phi}\bigg|_{\phi_c}\nabla_\mu(F^{\mu\nu J }\de\phi) \ ,
\ee
where 
\be
F_\de^{\mu\nu J}= \p^\mu\de A^\nu - \p^\nu\de A^\mu \ ,
\ee
and $F^{\mu\nu J}$ is the field strength corresponding to the attractor solution. We can simplify
\eqref{gaugefluc} using the attractor equation for the gauge field \eqref{gfeq}, where the
Chern-Simons term vanishes and the scalars are independent of spacetime coordinates at
the attractor point. Thus we have
\be\label{gaugefluc2}
a_{IJ}|_{\phi_c}\nabla_\mu F^{\mu\nu J}_{\de}=-\frac{\p
a_{IJ}}{\p\phi}\bigg|_{\phi_c}F^{\mu\nu J} \p_\mu\de\phi \ .
\ee
For the gauge field ansatz \eqref{gaugeansatz}, the non-trivial field strength component is only
along the $F^{\hx\hy}$ direction. Since the scalar field fluctuation in \eqref{linearisedfluc}
depends only on the radial direction, the right hand side of \eqref{gaugefluc2} vanishes. Hence,
there are no gauge field fluctuations that are sourced by the scalar fluctuations in this case. Thus
the linearized fluctuations of the gauge field about the attractor value satisfy the attractor
equation
\be
a_{IJ}|_{\phi_c}\nabla_\mu F^{\mu\nu J}_{\de}=0 \ .
\ee
From the point of view of the attractor mechanism in supergravity
\cite{Ferrara:1995ih,Strominger:1996kf}, it is the behavior of the scalar fields that is most
relevant for our case. Hence, we do not consider any independent gauge field fluctuations here.
Thus, we can drop the gauge field fluctuations for the rest of the
analysis in the following sections. 

In a general situation as opposed to the simple example considered here, the source term in
\eqref{gaugefluc2} need not vanish. In such a case, however one may still be able to solve the
problem in certain situations where the scalar fluctuation equations decouple from gauge field
fluctuations at linearized level \cite{Inbasekar:2013vra}. So solving the linearized equation for
scalar fluctuations determines the source term in the gauge field fluctuation, which can then in
principle be solved. However, the situation becomes more complicated for the metric fluctuations
since both the gauge field and scalar fluctuations will enter through the stress tensor.

Another notable simplification is that currently we are working with the $U(1)_R$ gauged
supergravity. When the gauging of the symmetries of scalar manifold is also considered there
are additional terms in \eqref{gaugefluc} and solving for the gauge field fluctuations is much
harder in the presence of additional scalar source terms.\footnote{See for example, eq 3.5 of
\cite{Inbasekar:2013vra}.} 

\subsection{Scalar fluctuations}\label{scalarflucanalysis}
 We will now solve the linearized equations for the scalar fluctuations about the attractor value
$\phi_c$. The linearized equation for the scalar field obtained from \eqref{lagrangian} takes a
remarkably simple form,
\be\label{scalarfluc}
g(\phi_c)\nabla_\mu\nabla^\mu\de\phi-\frac{\p^2\V_{attr}}{\p\phi^2}\bigg|_{\phi_c}\de\phi=0 ,
\ee
where $g(\phi)$ and the attractor potential are defined in \eqref{scalarmetric} and
\eqref{attractorpot} respectively. Using \eqref{hessianeigenvalue}, we define
\be\label{lambda}
\la=\frac{1}{g(\phi_c)}\frac{\p^2\V_{attr}}{\p\phi^2}\bigg|_{\phi_c}=\frac{-7+8\beta_t^2}{3} 
\ee
which is positive for the solution of interest, since $\beta_t^2>\frac{7}{8}$. Using the expression
for the metric \eqref{typeIIIsoln}, equation \eqref{scalarfluc} can be simplified as
\be
\big[\hr^2\p^2_{\hr}+(1+2\beta_t)\hr\p_{\hr}-\la\big]\de\phi=0 \ .
\ee
The general solution for this equation is of the form
\be
\de\phi= C_1 \hr^{\sqrt{\la+\beta_t^2}-\beta_t}+C_2 \hr^{-\sqrt{\la+\beta_t^2}-\beta_t} \ .
\ee
The type III metric \eqref{typeIII} is written in a coordinate system such that the horizon is
located at $\hr=0$. We require the scalar fluctuations \eqref{linearisedfluc} to vanish as
$\hr^\De$ for $\De>0$ such that the scalar field approaches its attractor value as $\hr\rightarrow
0$. Therefore, we choose $C_2=0$. The other constant $C_1$ cannot be fixed at this stage as the
equation \eqref{scalarfluc} is valid only near the horizon. However, we can choose $C_1=C_s \in
\mathbb{R}$ since the scalar fields in five dimensional gauged supergravity are real. In addition,
for non-trivial fluctuations $C_s\neq0$. Thus the scalar fluctuations which are well behaved near
the horizon are of the form
\be\label{scalarflucfinal}
\de\phi=C_s \hr^\De \ , \quad \De=\sqrt{\la+\beta_t^2}-\beta_t\ .
\ee
Note that, the condition obtained from \eqref{hessianeigenvalue} indeed ensures that the
scalar fluctuations are well behaved as $\hr\rightarrow 0$ near the horizon. 

To fix the constants in the solution completely, one has to solve the scalar equation in the
background of a solution which interpolates from Bianchi III to $AdS$ with appropriate boundary
conditions. Such interpolating metrics obeying reasonable energy conditions that interpolate to
Lifshitz or $AdS_2\times S^3$ which can then be connected to $AdS$ have been constructed
numerically in \cite{Kachru:2013voa}. However, they are not yet known to arise as solutions
to Einstein gravity coupled to some simple matter theory.

\subsection{Metric fluctuations}\label{metricflucanalysis}
In this section, we solve the linearized metric fluctuations about the type III metric,
that are sourced by scalar fluctuations \eqref{scalarflucfinal}. The linearized fluctuation
equations of the metric have the form \cite{Inbasekar:2013vra},
\begin{align}\label{linearisedmetriceq}
 \nabla^\al \nabla_\al \bar{\ga}_{\mu\nu}+ 2 R_{(\mu \ \nu) \ }^{\ \ \al \
\ \beta}\bar{\ga}_{\beta\al} - 2 R_{(\mu}^{ \ \ \beta} \bar{\ga}_{\nu)\beta}+g_{\mu\nu}
(R_{\al\beta}\bar{\ga}^{\al\beta}-\frac{2}{3}R \bar{\ga})+ R\bar{\ga}_{\mu\nu} \nn\\ + 2
(\dot{T}_{\mu\nu}^{attr}(g_{\al\beta}+\ep\ga_{\al\bet})|_{\ep=0}+\dot{T}_{\mu\nu}^{attr}
(\phi_c+\ep\de\phi)|_{\ep=0})=0 ,
\end{align}
where 
\be
\bar{\ga}_{\mu\nu}=\ga_{\mu\nu}-\frac{1}{2}\ga g_{\mu\nu}\ , \quad \ga=g^{\mu\nu}\ga_{\mu\nu} \ ,
\quad \bar{\ga}=-\frac{3}{2}\ga \ .
\ee
The dots indicate derivatives with respect to $\ep$. The covariant derivatives, raising and
lowering are with respect to the near horizon metric $g_{\mu\nu}$. The Riemann tensor, Ricci
tensor and curvature that appear in \eqref{linearisedmetriceq} are also with respect to
$g_{\mu\nu}$.

The contribution of the linearized metric fluctuations from the stress energy tensor are
\begin{align}\label{stressenergymetricdependence}
\dot{T}_{\mu\nu}^{attr}(g_{\al\beta}+\ep\ga_{\al\bet})|_{\ep=0}=& \V_{attr}|_{\phi_c}
(\bar{\ga}_{\mu\nu}-\frac{2\bar{\ga}}{3}g_{\mu\nu})\nn\\
&-(\bar{\ga}_{\la\si}-\frac{\bar{\ga}}{3} g_ { \la\si})
(\frac{1}{2}T_{attr}^{\la\si}|_{\phi_c}g_{\mu\nu}+a_{IJ}|_{\phi_c} F^{I \ \la}_{\mu}F^{J \
\si}_{\nu}).
\end{align}
where 
\be
T_{\mu\nu}^{attr}=\V_{attr}|_{\phi_c}g_{\mu\nu}-a_{IJ}|_{\phi_c}F_{\mu\la}^I F_\nu^{\ \la J} 
\ee
and $V_{attr}|_{\phi_c}$ is defined by \eqref{attrpotatcrit}. The contribution of the linearized
scalar fluctuations from the stress energy tensor are
\be
\dot{T}_{\mu\nu}^{attr}(\phi_c+\ep\de\phi)|_{\ep=0}=
\frac{\p \V_{attr}}{\p\phi}\bigg|_{\phi_c}g_{\mu\nu}\de\phi-\frac{\p a_{IJ}}{\p\phi}\bigg|_{\phi_c}
F^I_{\mu\la}F_{\nu}^{\ \la J} \de\phi \ ,
\ee
which can be further simplified using the attractor equation \eqref{scattr} to get
\be\label{stressenergyscalardependence}
\dot{T}_{\mu\nu}^{attr}(\phi_c+\ep\de\phi)|_{\ep=0}=-\frac{\p a_{IJ}}{\p\phi}\bigg|_{\phi_c}
F^I_{\mu\la}F_{\nu}^{\ \la J} \de\phi \ .
\ee

We can now solve for the metric fluctuations by plugging in the scalar fluctuations
\eqref{scalarflucfinal}. First, let us simplify the form of \eqref{linearisedmetriceq} by making a
few observations. We note that the type III metric in its explicit form 
\be
ds^2= -\hr^{2 \beta_t}d\hht^2+ \frac{d \hr^2}{\hr^2}+ d\hx^2+ e^{-2\hx}d\hy^2+
{\hr}^{2\beta_t}d\hz^2
\ee
is diagonal. Therefore, It is reasonable to expect fluctuations only along the diagonal directions.
Hence we can choose the fluctuations $\ga_{\mu\nu}$ to be symmetric. As a result the antisymmetrized
terms in \eqref{linearisedmetriceq} vanish, as can be checked explicitly. Thus we have
\begin{align}\label{linearisedmetriceqfinal}
 \nabla^\al \nabla_\al \bar{\ga}_{\mu\nu}+g_{\mu\nu}
(R_{\al\beta}\bar{\ga}^{\al\beta}-\frac{2}{3}R \bar{\ga})+ R\bar{\ga}_{\mu\nu}+& 2
(\dot{T}_{\mu\nu}^{attr}(g_{\al\beta}+\ep\ga_{\al\bet})|_{\ep=0} \nn\\&+\dot{T}_{\mu\nu}^{attr}
(\phi_c+\ep\de\phi)|_{\ep=0})=0 ,
\end{align}
with the contributions from the stress energy tensor corresponding to metric and scalar
fluctuations as given by \eqref{stressenergymetricdependence}
and \eqref{stressenergyscalardependence} respectively.

We choose the fluctuation terms of the metric in $g_{\mu\nu}+\ep\ga_{\mu\nu}(\hr)$ to be of the form
\begin{align}\label{metricpert}
&\ga_{\hht\hht}= C_{\hht} \hr^{2 \beta_t}  \gat_{\hht\hht}(\hr) \ , \nn\\
&\ga_{\hr\hr}=C_{\hr} \frac{1}{\hr^2}\gat_{\hr\hr}(\hr) \ ,\nn \\
&\ga_{\hx\hx}=C_{\hx} \gat_{\hx\hx}(\hr) \ ,\nn \\
&\ga_{\hy\hy}=C_{\hy} e^{-2\hx} \gat_{\hy\hy}(\hr) \ ,\nn\\
&\ga_{\hz\hz}=C_{\hz} \hr^{2\beta_t}\gat_{\hz\hz}(\hr) \ ,
\end{align}
where $C_{\hht},C_{\hr},C_{\hx},C_{\hy},C_{\hz}$ are constants which are to be determined in terms
of the gauged supergravity parameters $g_R, V_0,V_1,$ and the coefficient $C_s$ in the scalar
fluctuation \eqref{scalarflucfinal}. 

Because of the way the perturbations have been chosen in \eqref{metricpert}, one can contract the
Einstein equations with the vielbeins and write the final expressions in terms of the
$\gat_{\mu\nu}(\hr)$. We also observe that the source term from the scalar fluctuation
\eqref{stressenergyscalardependence} appears only in the $\hx\hx$ and $\hy\hy$ directions. While the
source goes like $\hr^\De$, the Einstein equations will contain terms like $\hr^2
\p_{\hr}^2\gat_{\mu\nu}$ , $\hr \p_{\hr} \gat_{\mu\nu}$ , $\gat_{\mu\nu}$. Hence one expects the
fluctuations $\gat_{\mu\nu}$ to also go like $\hr^\De$. This can be checked by observing
the explicit equations, which are rather messy.  We refer the reader to the appendix
\S\ref{linearisedeeq} for more details. Thus all the metric fluctuations should have the behavior
\be\label{metricflucsol}
\gat_{\hht\hht}=\gat_{\hr\hr}=\gat_{\hx\hx}=\gat_{\hy\hy}=\gat_{\hz\hz}=\hr^\De \ .
\ee
We now substitute \eqref{metricflucsol} in eqs. \eqref{linearisedmetriceqfinal} and reduce them to
an algebraic system,
\begin{align}
&\begin{split}\label{algebraic}
4 ({\beta_t}^2 (3 {C_{\hr}}&+3 {C_{\hht}}+{C_{\hx}}+{C_{\hy}}+3 {C_{\hz}})+2
{C_{\hht}}+{C_{\hx}}+{C_{\hy}})\\&+6 {\beta_t} \Delta 
({C_{\hr}}-{C_{\hht}}+{C_{\hx}}+{C_{\hy}}+{C_{\hz}})
+\Delta ^2({C_{\hr}}-{C_{\hht}}+{C_{\hx}}+{C_{\hy}}+{C_{\hz}})=0 \ , \nn
\end{split}
\end{align}
\begin{align}
&\begin{split}
 {C_{\hr}} (-4 (5 {\beta_t}^2+{\beta_t}+&1) +2 ({\beta_t}-2) \Delta +\Delta ^2)-2 ({\beta_t}-2)
\Delta
 ({C_{\hht}}+{C_{\hx}}+{C_{\hy}}+{C_{\hz}})\\&+4 {\beta_t} ({\beta_t}
(-{C_{\hht}}+{C_{\hx}}+{C_{\hy}}-{C_{\hz}})+{C_{\hht}}+{C_{\hx}}+{C_{\hy}}+{C_{\hz}})\\&+\Delta ^2
(-({C_{\hht}}+{C_{\hx}}+{C_{\hy}}+{C_{\hz}}))-4 ({C_{\hht}}+2
({C_{\hx}}+{C_{\hy}})+{C_{\hz}})=0 \ ,\nn
\end{split}
\end{align}
\begin{align}
&\begin{split}
 (16-32 {\beta_t}^2) {C_s}-{\phi_c} ((4 {\beta_t}^2+2 {\beta_t} \Delta +\Delta ^2)
({C_{\hr}}&+{C_{\hht}}+{C_{\hy}}+{C_{\hz}})\nn\\&+{C_{\hx}} (12 {\beta_t}^2-2 {\beta_t} \Delta
-\Delta
^2+12))=0  \ , \nn
\end{split}
\end{align}
\begin{align}
&\begin{split}
(16-32 {\beta_t}^2) {C_s}&-{\phi_c} \big(4 {\beta_t}^2 ({C_{\hr}}+{C_{\hht}}+{C_{\hx}}+3
{C_{\hy}}+{C_{\hz}})+2 {\beta_t} \Delta 
({C_{\hr}}+{C_{\hht}}+{C_{\hx}}-{C_{\hy}}+{C_{\hz}})
\\&+\Delta ^2({C_{\hr}}+{C_{\hht}}+{C_{\hx}}-{C_{\hy}}+{C_{\hz}})+6
({C_{\hr}}+{C_{\hht}}+{C_{\hx}}+{C_{\hy}}+{C_{\hz}})\big) =0 \ , \nn
\end{split}
\end{align}
\begin{align}
-4 {\beta_t}^2 (3 {C_{\hr}}+3 {C_{\hht}}+{C_{\hx}}+&{C_{\hy}}+3 {C_{\hz}})-6 {\beta_t} \Delta 
({C_{\hr}}+{C_{\hht}}+{C_{\hx}}+{C_{\hy}}-{C_{\hz}})\nn\\ &-\Delta ^2
({C_{\hr}}+{C_{\hht}}+{C_{\hx}}+{C_{\hy}}-{C_{\hz}})-4 ({C_{\hx}}+{C_{\hy}}+2 {C_{\hz}})=0 \ ,
\end{align}
which can be solved to determine the coefficients. Note that the other parameters $\phi_c,\De,
\beta_t$ that enter the equations are all expressible in terms of the gauged supergravity parameters
$g_R,V_0,V_1$ from eqs \eqref{critpoint} and \eqref{scalarflucfinal}. However, we will express
everything in terms of $\beta_t$ for convenience. Thus the solution for the coefficients are,
\begin{align}\label{solcoeff}
&C_{\hht}=\frac{C_s}{\phi_c} F_0(\beta_t)\ ,\nn\\
&C_{\hr}=\frac{C_s}{\phi_c} F_1(\beta_t)\ ,\nn\\
&C_{\hx}=\frac{C_s}{\phi_c} F_2(\beta_t)\ ,\nn\\
&C_{\hy}=\frac{C_s}{\phi_c} F_3(\beta_t)\ ,\nn\\
&C_{\hz}=\frac{C_s}{\phi_c} F_4(\beta_t)\ .
\end{align}
where $F_i(\beta_t), i=0,\ldots4$ are complicated functions of $\beta_t$ which are given in
\S\ref{funcofbetat}. Note that all the coefficients are proportional to the coefficient $C_s$. This
is a consistency check that the metric fluctuations considered in the analysis are sourced by the
scalar fluctuations.

Thus the full metric along with the fluctuations is
\begin{align}\label{metricfinal}
 ds^2=
-\bigg(1+C_{\hht}\hr^\De\bigg)\hr^{2\beta_t}d\hht^2&+\bigg(1+C_{\hr}\hr^\De\bigg)\frac{d\hr^2}{\hr^2
}+\bigg(1+C_{ \hx} \hr^\De\bigg)d\hx^2\nn\\
&+\bigg(1+C_{\hy}\hr^\De\bigg)e^{-2\hx}d\hy^2+\bigg(1+C_{\hr}\hr^\De\bigg)\hr^{2\beta_t}d\hz^2 \ .
\end{align}
From eq \eqref{lambda} and eq \eqref{scalarflucfinal}, we see that positivity of $\la$ implies $\De$
is positive for the solution \eqref{typeIIIsol}. Hence, all the metric fluctuations are well behaved
and the metric approaches the type III attractor metric as one approaches the horizon
$\hr\rightarrow 0$. The reader may worry that the perturbation in $\hr\hr$ is well
behaved only if $\De>2$. However there is no need to put any additional condition, since
the behavior at $\hr\rightarrow 0$ is dictated by the $\frac{1}{\hr^2}$ term owing to $\De$
being positive. Thus we have constructed a stable Bianchi III attractor solution in gauged
supergravity. In the following section, we investigate the supersymmetry of this solution.

\section{Supersymmetry analysis}\label{susy}
In this section, we analyze the Killing spinor equations for the $U(1)_R$ gauged supergravity with
the Bianchi type III solution \eqref{typeIIIsol} as the background. The Killing spinor equation is
obtained by setting the supersymmetric variation of the gravitino to zero. For the $\N=2 ,
U(1)_R$ gauged supergravity the gravitino variation is 
\cite{Gunaydin:1999zx},
\be\label{killingspinor}
\de\psi_{\mu
i}=\nabla_\mu(\om)\ep_i+\frac{i}{4\sqrt{6}}h_I(\ga_{\mu\nu\rho}-4 g_{\mu\nu}\ga_\rho)F^{ I\nu\rho}
\ep_i+\de'\psi_{\mu i} \ .
\ee

Our notations and conventions are summarized in \S\ref{notations}. The indices $I$ label the number
of vectors and the scalars $h_I$ are as defined in \S\ref{gaugedsugra}. Although we have only one
gauge field for the solution \eqref{typeIIIsol}, we will keep the $I$ indices for the gauge fields
to avoid introducing the explicit form of $h_I$ in the equations. The term $\de'\psi_{\mu i}$ is
the modification in the supersymmetry variations as a result of the $U(1)_R$ gauging. Explicitly it
takes the form,
\be
\de'\psi_{\mu i}=-\frac{i}{\sqrt{6}}g_R h^I V_I \ga_\mu \de_{ij}\ep^j \ ,
\ee
where $V_I$ are the parameters that appear in the $U(1)_R$ gauging. Note that the $\de_{ij}$ is not
used to raise or lower the $SU(2)$ index. 

We now proceed to analyze the Killing spinor equations. The vielbeins and spin connections of the
metric \eqref{typeIIIsol} are 
\begin{align}
 &e^0_{\hht}=r^{\beta_t} \ , \ e_{\hr}^1=\frac{1}{\hr} \ , \ e_{\hx}^2 = 1 \ , \ e_{\hy}^3
=e^{-\hx} \ , \ e_{\hz}^4=\hr^{\beta_t} \ , \nn\\
 & \om^{01}_{ \ \ \hht}=\beta_t \hr^{\beta_t} \ , \ \om^{32}_{ \ \ \hy}=-e^{-\hx} \ , \
\om^{41}_{\ \ \hz}=\beta_t \hr^{\beta_t} \ .
\end{align}
Substituting the above in \eqref{killingspinor}, the Killing spinor equations can be written as
\begin{align}\label{killingspinor1}
 & \ga_0 \hr^{-\beta_t} \p_{\hht}\ep_i-\frac{\beta_t}{2}
\ga_1\ep_i+\frac{i}{2\sqrt{6}}A_3^Ih_I\ga_{23}\ep_i+\frac{i}{\sqrt{6}}g_Rh^IV_I\de_{ij}\ep^j=0 \ ,
\nn\\
 & \ga_1 \hr
\p_{\hr}\ep_i-\frac{i}{2\sqrt{6}}A_3^Ih_I\ga_{23}\ep_i-\frac{i}{\sqrt{6}}g_Rh^IV_I\de_{ij}\ep^j=0 \
, \nn\\
& \ga_2 \p_{\hx}\ep_i+\frac{i}{\sqrt{6}}A_3^Ih_I
\ga_{23}\ep_i-\frac{i}{\sqrt{6}}g_Rh^IV_I\de_{ij}\ep^j=0 \ , \nn\\
& \ga_3 e^{\hx}\p_{\hy}\ep_i-\frac{1}{2}\ga_2\ep_i+\frac{i}{\sqrt{6}}A_3^I
h_I\ga_{23}\ep_i-\frac{i}{\sqrt{6}}g_Rh^IV_I\de_{ij}\ep^j=0 \ , \nn\\
& \ga_4 \hr^{-\beta_t}\p_{\hz}\ep_i
+\frac{\beta_t}{2}\ga_1\ep_i-\frac{i}{2\sqrt{6}}A_3^Ih_I\ga_{23}\ep_i-\frac{i}{\sqrt{6}}
g_Rh^IV_I\de_{ij}\ep^j=0 \ .
\end{align}
The $\ga_a$ matrices that appear in the above set of equations are in tangent space.

We choose a radial profile for the Killing spinor. This is motivated by the fact
that the radial spinor preserves the time translation and homogeneous symmetries of the
type III metric \eqref{typeIII}. Moreover, it is well known that the radially dependent spinor
generates the Poincar\'{e} supersymmetries in $AdS$ \cite{Lu:1996rhb,Lu:1998nu}. Furthermore, some
of the Bianchi type I solutions such as the Lifshitz and $AdS_3\times\mathbb{R}^2$ solutions in
gauged supergravity preserve 1/4 of the supersymmetries for the radial spinor
\cite{Cassani:2011sv,Halmagyi:2011xh,Almuhairi:2011ws}.

We choose the spinor ansatz
\be\label{ansatz}
\ep_i= f(\hr) \chi_i \ ,
\ee
where $\chi_i$ is a constant symplectic majorana spinor. Substituting \eqref{ansatz} in the Killing
spinor equation \eqref{killingspinor1}, we see that $\hht,\hz$ equations become identical. Adding
the $\hht$ equation and the radial equation we get
\be
\hr\p_{\hr}f(\hr)-\frac{\beta_t}{2}f(\hr)=0 \ ,
\ee
which is solved by 
\be
f(\hr)=\hr^{\frac{\beta_t}{2}} .
\ee
Using the above in \eqref{ansatz} and substituting it in the Killing spinor
equation \eqref{killingspinor1} we get,
\begin{align}
 & \frac{\beta_t}{2}
\ga_1\chi_i-\frac{i}{2\sqrt{6}}A_3^Ih_I\ga_{23}\chi_i-\frac{i}{\sqrt{6}}g_Rh^IV_I\de_{ij}\chi^j=0 \
,
\nn\\
& \frac{i}{\sqrt{6}}A_3^Ih_I \ga_{23}\chi_i-\frac{i}{\sqrt{6}}g_Rh^IV_I\de_{ij}\chi^j=0 \ , \nn\\
& \frac{1}{2}\ga_2\chi_i-\frac{i}{\sqrt{6}}A_3^I
h_I\ga_{23}\chi_i+\frac{i}{\sqrt{6}}g_Rh^IV_I\de_{ij}\chi^j=0 \ .
\end{align}
From the last two of the above equations, it follows that 
\be
\ga_2\chi_i=0 \ .
\ee
This condition breaks all of the supersymmetry. The origin of the $\ga_2$ term is the spin
connection term due to the $EAdS_2$ \eqref{AdS2EAdS2} part of the type III metric. Thus, a naive
radial spinor does not preserve supersymmetry in this case. This suggests that the
stable Bianchi III metric we have constructed may be a non-supersymmetric attractor. However, it is
possible that there may be a more general ansatz similar to the one studied in \cite{Klemm:2000nj}
for a black string solution that interpolates between $AdS_3\times \mathbb{H}^2$ and
$AdS_5$ in a $U(1)^3$ gauged supergravity. We hope to explore this in detail in future works.

\section{Summary and conclusions}\label{summary}
We studied the $AdS_3\times \mathbb{H}^2$ solution which is a special case of the Bianchi type III
class in $U(1)_R$ gauged supergravity. We found that there exist a class of such solutions
parametrized by $g_R,V_0,V_1$ that satisfied the two sufficient requirements for the attractor
mechanism, namely the existence of a critical point of the attractor potential and that the Hessian
of the attractor potential should have a positive eigenvalue. 

We investigated the stability of the solution in gauged supergravity by studying
the linearized fluctuations of the gauge field, scalar field, metric about their attractor values.
The stress energy tensor in gauged supergravity depends on linearized fluctuations of scalars
and gauge fields \cite{Inbasekar:2013vra}. In order to avoid backreaction and deviation from the
attractor geometry, all the fluctuations have to be well behaved as one approaches the horizon.

For the solution \eqref{typeIIIsol}, we showed that the source term in the gauge field fluctuations
vanishes. Thus there are no gauge field fluctuations sourced by scalar fluctuations. The metric
fluctuation equations are sourced completely by the scalar perturbations. We showed that for the
solution satisfying the sufficient conditions for the attractor mechanism, the scalar
fluctuations are well behaved near the horizon. We also solved the metric fluctuations and showed
that all the fluctuations are regular. Since all the linearized fluctuations are well behaved near
the horizon, we infer that the type III Bianchi solution is a stable attractor solution at the
linearized level.

One of the simplifications that aided us in the stability analysis was that there were no gauge
field fluctuations which are sourced by scalar fluctuations. As we commented before in
\S\ref{gaugeflucanalysis}, this need not happen in general. For more complicated situations we
expect that as long as the solution satisfies the sufficient conditions for the attractor mechanism
\cite{Goldstein:2005hq}, the Bianchi type geometries might be stable with respect to linearized
fluctuations about the attractor values. We hope to explore these aspects and
look for more interesting solutions in future.

In the long run, we hope our stability analysis will provide motivation to explore the possibility
of construction of analytic black brane solutions which interpolate between IR and UV attractor
geometries. In particular, it will be very interesting to construct solutions that are
asymptotically $AdS$. Such interpolating solutions will be helpful to explore the
holographic duals of Bianchi attractors. Recent progress in this direction include numerical
solutions which interpolate between Bianchi types and Lifshitz or $AdS_2\times S^3$ from where they
can be connected to anti de-Sitter space \cite{Kachru:2013voa}. It will be valuable to construct
analytic interpolating solutions in a simple theory of gravity coupled to suitable matter.

In this paper, we also investigated the supersymmetry of the Bianchi type III solution. We studied
the Killing spinor equations of $\N=2, U(1)_R$ gauged supergravity with the background metric
\eqref{typeIIIsol}. We chose a radial profile for the Killing spinor since it preserves the time
translations and homogeneous symmetries of the metric. However, we found that the naive radial
spinor which gives supersymmetric Bianchi I spaces such as $AdS$ and Lifshitz fails for this
case. This suggests that the stable type III solution we obtained may be a non-supersymmetric
attractor. It would be interesting to construct supersymmetric Bianchi attractors in gauged
supergravity along the lines of the $AdS_3 \times \mathbb{H}^2$ solution in \cite{Klemm:2000nj}. In
a related exploration, it would be worthwhile to construct Bianchi attractors from wrapped branes
\cite{Maldacena:2000mw} in supergravity. We hope to report these in future works.

\acknowledgments
We would like to thank Sandip Trivedi for motivation and stimulating discussions throughout the
course of this work. We would like to thank Prasanta Tripathy and Sandip Trivedi for comments on
earlier versions of the draft. We would also like to thank Arpan Saha, Prasanta Tripathy and Sandip
Trivedi for collaboration in a related work. K.I would like to thank Sachin Jain and Shuichi
Yokoyama for discussions. Most of all we would like to thank the people of India for
generously supporting research in string theory.
\appendix
\section{Notations and conventions}\label{notations}
In this section, we summarize our notations and conventions on tangent space and spinors. We use
greek indices for spacetime and roman for tangent space. Our conventions for the flat tangent space
metric is $\eta_{ab}=(-,+,+,+,+)$. The tangent space indices are denoted by $a,b=0,1,2,3,4$. 

The tangent space matrices satisfy the usual Clifford algebra
\be
\{ \ga_a,\ga_b\}=2\eta_{ab}\ .
\ee
Antisymmetrization is done with the following convention,
\be
\ga_{a_1 a_2 \ldots a_n}= \ga_{[a_1 a_2 \ldots a_n]}= \frac{1}{n!}\sum_{\si \in P_n} Sign(\si)
\ga_{a_{\si(1)}}\ga_{a_{\si(2)}}\ldots \ga_{a_{\si(n)}} \ .
\ee
In $d=5$ only $I, \ga_a , \ga_{ab}$ form an independent set, other matrices are related by the
general identity for $d=2k+3$,
\be
\ga^{\mu_1\mu_2\ldots\mu_s}=
\frac{-i^{-k+s(s-1)}}{(d-s)!}\ep^{\mu_1\mu_2\ldots\mu_s}\ga_{\mu_{s+1}\ldots\mu_d} \ .
\ee
We also recollect that the spinors in five dimensions satisfy the symplectic majorana condition
\be
\bar{\ep}^i\equiv (\ep_i^*)^t\ga^0=(\ep^i)^t C \ ,
\ee
where $C$ is the charge conjugation matrix which obeys $C^t=C^{-1}=-C$. 

Unlike the case in four dimensions, the $SU(2)$ indices are not raised and lowered by complex
conjugation. Instead they are raised and lowered by the $SU(2)$ covariant tensor with the
conventions $\ve_{12}=\ve^{12}=1$. Note that the $SU(2)$ indices are always raised or lowered in the
NW-SE direction
\be
\ep^i=\ve^{ij} \ep_j \ , \quad \ep_i=\ep^j \ve_{ji} \ .
\ee

The covariant derivative acting on $\ep_i$ is with respect to the Lorentz covariant spin
connection $\om_\mu^{ab}$ defined as
\be
\nabla_\mu(\om)\ep_i=\p_\mu\ep_i+\frac{1}{4}\om_\mu^{ab}\ga_{ab}
\ee
\section{Linearized Einstein equations}\label{linearisedeeq}
In this section, we provide the explicit form of the linearized equations that follow from
\eqref{linearisedmetriceqfinal}. We substitute the expressions for the attractor potential
\eqref{attrpotatcrit}, the scalar fluctuations \eqref{scalarflucfinal}, the terms from the stress
energy tensor \eqref{stressenergymetricdependence}, \eqref{stressenergyscalardependence} and the
metric fluctuations \eqref{metricpert} into the linearized Einstein equation
\eqref{linearisedmetriceqfinal}. We then contract it with the vielbeins $e^\mu_a$ to obtain the
following equations. The $\hht\hht$ equation is
\begin{align}
 \hr^2 {\gat_{\hr\hr}}''-\hr^2 {\gat_{\hht\hht}}''+\hr^2 {\gat_{\hx\hx}}''&+\hr^2
{\gat_{\hy\hy}}''+\hr^2 {\gat_{\hz\hz}}''+12 {\beta_t}^2 {\gat_{\hr\hr}}+4 (3 {\beta_t}^2+2)
{\gat_{\hht\hht}}+4 {\beta_t}^2 {\gat_{\hx\hx}}+4 {\beta_t}^2 {\gat_{\hy\hy}}
\nn\\&
+12 {\beta_t}^2{\gat_{\hz\hz}}+6 {\beta_t} \hr {\gat_{\hr\hr}}'-6 {\beta_t} \hr
{\gat_{\hht\hht}}'+6 {\beta_t} \hr{\gat_{\hx\hx}}'+6 {\beta_t} \hr {\gat_{\hy\hy}}'+6 {\beta_t}
\hr{\gat_{\hz\hz}}'\nn\\
&+\hr{\gat_{\hr\hr}}'-\hr {\gat_{\hht\hht}}'+\hr {\gat_{\hx\hx}}'
 +4 ({\gat_{\hx\hx}}+{\gat_{\hy\hy}})+\hr{\gat_{\hy\hy}}'+\hr
{\gat_{\hz\hz}}'=0 \ .
\end{align}
The $\hr\hr$ equation is
\begin{align}
\hr^2 {\gat_{\hr\hr}}''-\hr^2 {\gat_{\hht\hht}}''&-\hr^2 {\gat_{\hx\hx}}''-\hr^2 
{\gat_{\hy\hy}}''-\hr^2 {\gat_{\hz\hz}}''-4 (5 {\beta_t}^2+{\beta_t}+1) {\gat_{\hr\hr}}+4
{\beta_t}^2 {\gat_{\hx\hx}}+4 {\beta_t}^2 {\gat_{\hy\hy}}\nn \\
&-4 {\beta_t}^2{\gat_{\hz\hz}}+2{\beta_t}\hr {\gat_{\hr\hr}}'-2 {\beta_t} \hr {\gat_{\hht\hht}}'-4
({\beta_t}-1) {\beta_t}
{\gat_{\hht\hht}}-2 {\beta_t} \hr {\gat_{\hx\hx}}'+4 {\beta_t} {\gat_{\hx\hx}}\nn\\
&-2 {\beta_t} \hr
{\gat_{\hy\hy}}'+4 {\beta_t} {\gat_{\hy\hy}}-2 {\beta_t} \hr {\gat_{\hz\hz}}'+4 {\beta_t}
{\gat_{\hz\hz}}-3 \hr {\gat_{\hr\hr}}'+3 \hr {\gat_{\hht\hht}}'\nn\\
&-4 ({\gat_{\hht\hht}}+2
({\gat_{\hx\hx}}+{\gat_{\hy\hy}})+{\gat_{\hz\hz}})+3 \hr {\gat_{\hx\hx}}'+3 \hr
{\gat_{\hy\hy}}'+3\hr {\gat_{\hz\hz}}'=0 \ .
\end{align}
The $\hx\hx$ equation is
\begin{align}
 -\frac{(2 {\beta_t}^2-1) (8 {C_s} \hr^{\De }+{\phi_c} {\gat_{\hy\hy}})}{{\phi_c}}&-2 {\beta_t}^2
({\gat_{\hr\hr}}+{\gat_{\hht\hht}}+3 {\gat_{\hx\hx}}+{\gat_{\hz\hz}})-\frac{1}{2} \hr \big((2
{\beta_t}+1) {\gat_{\hr\hr}}'
\nn\\&+2 {\beta_t}({\gat_{\hht\hht}}'-{\gat_{\hx\hx}}'+{\gat_{\hy\hy}}'+{\gat_{\hz\hz}}')+\hr
({\gat_{\hr\hr}}''+{\gat_{\hht\hht}}''-{\gat_{\hx\hx}}''+{\gat_{\hy\hy}}''+{\gat_{\hz\hz}}'')\nn\\&+
{
\gat_{\hht\hht}}'-{\gat_{\hx\hx}}'+{\gat_{\hy\hy}}'+{\gat_{\hz\hz}}'\big)-6{\gat_{\hx\hx}}-{\gat_{
\hy\hy}}=0 \ .
\end{align}
The $\hy\hy$ equation is
\begin{align}
-\frac{16 (2 {\beta_t}^2-1) {C_s} \hr^{\De }}{{\phi_c}}&+2 (-2 {\beta_t}^2
({\gat_{\hr\hr}}+{\gat_{\hht\hht}}+3 {\gat_{\hy\hy}}+{\gat_{\hz\hz}})-{\gat_{\hx\hx}}-3
{\gat_{\hy\hy}})+2 (1-2 {\beta_t}^2) {\gat_{\hx\hx}}\nn\\
&-\hr\big((2{\beta_t}+1){\gat_{\hr\hr}}'+2{\beta_t}({\gat_{\hht\hht}}'+{\gat_{\hx\hx}}'-{\gat_{
\hy\hy } } '+
{\gat_{\hz\hz}}')+\hr({\gat_{\hr\hr}}''+{\gat_{\hht\hht}}''+{\gat_{\hx\hx}}''\nn\\&-{\gat_{\hy\hy}}
''+ {\gat_{ \hz\hz}}'')+ {\gat_
{\hht\hht}}'+{\gat_{\hx\hx}}'-{\gat_{\hy\hy}}'+{\gat_{\hz\hz}}'\big)-6 {\gat_{\hr\hr}}-6
{\gat_{\hht\hht}}-6 {\gat_{\hx\hx}}-6 {\gat_{\hz\hz}}=0 \ .
\end{align}
The $\hz\hz$ equation is
\begin{align}
\hr^2 (-{\gat_{\hr\hr}}'')-\hr^2 {\gat_{\hht\hht}}''&-\hr^2 {\gat_{\hx\hx}}''-\hr^2
{\gat_{\hy\hy}}''+\hr^2 {\gat_{\hz\hz}}''-12 {\beta_t}^2 {\gat_{\hr\hr}}-12 {\beta_t}^2
{\gat_{\hht\hht}}-4 {\beta_t}^2 {\gat_{\hx\hx}}-4 {\beta_t}^2 {\gat_{\hy\hy}}\nn\\
&-12 {\beta_t}^2{\gat_{\hz\hz}}-6 {\beta_t} \hr {\gat_{\hr\hr}}'-6 {\beta_t} \hr
{\gat_{\hht\hht}}'-6{\beta_t} \hr{\gat_{\hx\hx}}'-6 {\beta_t} \hr {\gat_{\hy\hy}}'+6 {\beta_t} \hr
{\gat_{\hz\hz}}'-\hr{\gat_{\hr\hr}}'\nn\\&-\hr {\gat_{\hht\hht}}'-\hr
{\gat_{\hx\hx}}'-4({\gat_{\hx\hx}}+{\gat_{\hy\hy}}+2{\gat_{\hz\hz}})-\hr {\gat_{\hy\hy}}'+\hr
{\gat_{\hz\hz}}'=0 \ .
\end{align}
In the above equations, the prime indicates derivative with respect to $\hr$. We see that all the
double derivatives are multiplied by $\hr^2$, while the single derivatives are multiplied by
$\hr$. Now, the $\hx\hx$ and $\hy\hy$ equations contain the source term which goes like $\hr^\De$.
It is then clear that the metric fluctuations $\gat_{\mu\nu}$ all go like $\hr^\De$.
\section{Coefficients of the linearized fluctuations}\label{funcofbetat}
The various functions that appear in the coefficients \eqref{solcoeff} are
\begin{align}
& F_0(\beta_t)=-64  ({\beta_t}^2+4) (2 {\beta_t}^2-1)
\frac{N_{\hht}^1(\beta_t)+N_{\hht}^2(\beta_t)}{D_1(\beta_t)+D_2(\beta_t)+D_
3(\beta_t)+D_4(\beta_t) } \ , \\
& F_1(\beta_t)= 64  ({\beta_t}^2+4) (2
{\beta_t}^2-1)\frac{N_{\hr}^1(\beta_t)+N_{\hr}^2(\beta_t)}{D_1(\beta_t)+D_2(\beta_t)+D_
3(\beta_t)+D_4(\beta_t) } \ , \\
& F_2(\beta_t)=8  (2
{\beta_t}^2-1)\frac{N_{\hx}^1(\beta_t)+N_{\hx}^2(\beta_t)+N_{\hx}^3(\beta_t)}{
D_1(\beta_t)+D_2(\beta_t)+D_3(\beta_t)+D_4(\beta_t)} \ , \\
& F_3(\beta_t)=8  (2
{\beta_t}^2-1)\frac{N_{\hy}^1(\beta_t)+N_{\hy}^2(\beta_t)+N_{\hy}^3(\beta_t)}{
D_1(\beta_t)+D_2(\beta_t)+D_3(\beta_t)+D_4(\beta_t)} \ , \\
& F_4(\beta_t)=-64  ({\beta_t}^2+4) (2
{\beta_t}^2-1)\frac{N_{\hz}^1(\beta_t)+N_{\hz}^1(\beta_t)}{
D_1(\beta_t)+D_2(\beta_t)+D_3(\beta_t)+D_4(\beta_t)} \ ,
\end{align}
where,
\begin{align}
& N_{\hht}^1(\beta_t)= 272 {\beta_t}^4+80 (f(\beta_t)-1) {\beta_t}^2+4
(f(\beta_t)-84) {\beta_t} \ , \nn \\
& N_{\hht}^2(\beta_t)=-4 f(\beta_t)+16 (7 f(\beta_t)+33) {\beta_t}^3+107 \ , \nn\\
 & N_{\hr}^1(\beta_t) = 304 {\beta_t}^4+8 (14 f(\beta_t)-53) {\beta_t}^2+4 (5
f(\beta_t)+84) {\beta_t} \ ,\nn\\
 & N_{\hr}^2(\beta_t) =28 f(\beta_t)+16 (5 f(\beta_t)-33) {\beta_t}^3-179 \ ,\nn\\
& N_{\hx}^1(\beta_t)=  4928 {\beta_t}^6+4 (1000 f(\beta_t)+4821) {\beta_t}^2-4 (53
f(\beta_t)-924) {\beta_t} \ , \nn\\
& N_{\hx}^2(\beta_t)=644 f(\beta_t)-64 (5 f(\beta_t)-33) {\beta_t}^5+16 (68 f(\beta_t)+1419)
{\beta_t}^4 \ , \nn\\
& N_{\hx}^3(\beta_t)=-16 (166 f(\beta_t)+447) {\beta_t}^3+671 \ , \nn \\
& N_{\hy}^1(\beta_t)= 4928 {\beta_t}^6+4 (1216 f(\beta_t)+6009) {\beta_t}^2-4 (107
f(\beta_t)+3612) {\beta_t} \ , \nn \\
& N_{\hy}^2(\beta_t)=-4 f(\beta_t)-64 (5 f(\beta_t)-33) {\beta_t}^5+16 (68 f(\beta_t)+1689)
{\beta_t}^4 \ , \nn\\
& N_{\hy}^3(\beta_t)=(21360-64 f(\beta_t)) {\beta_t}^3+7745 \ ,\nn \\
& N_{\hz}^1(\beta_t)=(272 {\beta_t}^4+80 (f(\beta_t)-1) {\beta_t}^2+4
(f(\beta_t)-84) {\beta_t} \ ,\nn \\
& N_{\hz}^2(\beta_t)= -4 f(\beta_t)+16 (7 f(\beta_t)+33) {\beta_t}^3+107) \ ,\nn \\
& D_1(\beta_t)=-33024 {\beta_t}^8-8 (3910
f(\beta_t)+13839) {\beta_t}^2+4 (367 f(\beta_t)-1428) {\beta_t} \ , \nn\\
& D_2(\beta_t)=-3276 f(\beta_t)+256 (25
f(\beta_t)+99) {\beta_t}^7-128 (58 f(\beta_t)+1525) {\beta_t}^6 \ , \nn\\
& D_3(\beta_t)=192 (147 f(\beta_t)+400)
{\beta_t}^5-32 (1178 f(\beta_t)+8565) {\beta_t}^4 \ , \nn\\
& D_4(\beta_t)=48 (309 f(\beta_t)-1045) {\beta_t}^3-10445 \ , \nn\\
& f(\beta_t)=\sqrt{-21+33\beta_t^2} \ .
\end{align}

\providecommand{\href}[2]{#2}\begingroup\raggedright\endgroup
\end{document}